\documentclass[12pt]{article}
\usepackage[utf8]{inputenc}
\usepackage{amsmath, amssymb, amsfonts,epsfig,subfigure,bm,cite}
\usepackage{graphicx}
\usepackage{hyperref}
\usepackage{geometry}

\title{Development of an Inclusive Educational Platform Using Open Technologies and Machine Learning: A Case Study on Accessibility Enhancement.}
\author{
    Jimi Togni \\
    jimitogni@gmail.com, j226359@dac.unicamp.br \\
    University of Campinas - UNICAMP
}
\date{January 05, 2025}

\begin{document}

\maketitle

\section*{Abstract}
\label{sec:abs}
This study addresses the pressing challenge of educational inclusion for students with special needs by proposing and developing an inclusive educational platform. Integrating machine learning, natural language processing, and cross-platform interfaces, the platform features key functionalities such as speech recognition functionality to support voice commands and text generation via voice input; real-time object recognition using the YOLOv5 model, adapted for educational environments; Grapheme-to-Phoneme (G2P) conversion for Text-to-Speech systems using seq2seq models with attention, ensuring natural and fluent voice synthesis; and the development of a cross-platform mobile application in Flutter with on-device inference execution using TensorFlow Lite. The results demonstrated high accuracy, usability, and positive impact in educational scenarios, validating the proposal as an effective tool for educational inclusion. This project underscores the importance of open and accessible technologies in promoting inclusive and quality education. \\

\noindent{}{\bf\large Keywords --} machine learning, natural language processing, education, inclusive technologies. \


\section{Introduction}
\label{sec:intro}

Educational inclusion for individuals with special needs is among the most significant challenges faced by modern societies \cite{intro1}. Despite advancements in legislation and assistive technologies, a considerable number of students continue to encounter barriers to accessing quality education. These difficulties are further compounded by the lack of accessible and personalized tools tailored to diverse learning needs, including those related to visual, auditory, motor, or cognitive impairments \cite{intro2}.

The advent of machine learning and natural language processing (NLP) has opened new avenues for developing innovative solutions in inclusive education \cite{intro_ml_3}. Technologies such as text-to-speech (TTS) readers, automatic translators for sign language (Libras), and personalized interfaces have the potential to transform education by adapting content and teaching methods to meet the individual needs of students \cite{intro_ml}.

An additional aspect explored in this project is speech recognition to offer broader accessibility. We implemented a system based on the Whisper model \cite{whisper}, which not only converts speech to text but also distinguishes between control commands, such as "start reading" or "translate to Libras," and textual inputs for document creation. This feature enhances interactivity and usability, allowing users to control the system's features via voice commands or use voice input for text dictation \cite{speech_book}.

In this context, this project proposes the development of technological tools aimed at promoting educational inclusion, focusing on integrated solutions based on open technologies that leverage machine learning (ML) and natural language processing (NLP) \cite{intro_ml_2}. The primary objectives include creating an inclusive educational assistant capable of supporting voice commands and enabling text creation through speech input. Additionally, the project aims to provide a text-to-speech tool, sign language translation, and a convolutional neural network model to describe and recognize objects in real-time using mobile devices.

This project is justified by the necessity of making learning more equitable and accessible for all students, contributing to inclusive and high-quality education. Furthermore, it seeks to harness the potential of open technologies to create sustainable, replicable solutions adaptable to various educational contexts.


\section{Literature Review}
\label{sec:rev}

Virtual assistants powered by machine learning have become increasingly prominent in educational settings due to their ability to adapt content and teaching methods. According to \cite{ref1}, TTS technologies integrated with NLP can significantly enhance the engagement and performance of visually impaired students. Moreover, studies such as \cite{ref2} highlight the role of automatic translators for Libras in promoting the inclusion of deaf students.

Personalized assessments are essential to ensure equitable conditions for all students to demonstrate their knowledge. Studies like \cite{ref3} show how frameworks based on React and NLP models can be utilized to create interactive and adaptive assessments. The works in \cite{ref4,ref5} explored the use of eye-tracking to assist students with motor limitations and learning difficulties in completing activities.

Inclusive text editors with features like voice reading and Braille export have been explored in studies like \cite{ref6}. In \cite{ref7}, the authors presented a system that uses real-time dictation with SpeechRecognition, while \cite{ref8, ref9} developed an editor integrating OCR to convert documents into accessible formats.

The use of machine learning for educational diagnostics is widely recognized as an effective approach for identifying learning difficulties. \cite{ref10, ref11} implemented a system that evaluates students’ performance in real-time and suggests interventions based on regression and classification models, providing personalized reports for educators.

This literature review aims to highlight the relevance of the topics addressed in this project, emphasizing recent advancements and identifying gaps that justify the development of the proposed solutions.


\section{Methodology}
\label{sec:met}

The methodology employed in this project was designed to integrate machine learning, natural language processing (NLP), deep neural networks, and technological tools. It consists of stages that include research, development, validation, and implementation on cross-platform systems, aiming to create accessible, effective, and impactful educational solutions.

This project adopted an approach based on deep neural networks and the YOLOv5 model \cite{yolo} for object recognition, along with seq2seq models for grapheme-to-phoneme (G2P) conversion tasks \cite{g2p}. Additionally, a speech recognition system \cite{speech_book} was integrated, based on the Whisper model \cite{whisper}, to support voice interactions, offering features such as system control and text dictation.

The Whisper model \cite{whisper} was integrated into the system to process real-time speech input. The model was fine-tuned to differentiate voice commands from textual inputs, using a hybrid flow combining keyword detection and NLP-based semantic classification \cite{speech_2}. Through this approach, commands such as "start reading" are recognized and executed, while textual inputs are directly transcribed for editing or document creation.

To refine the ML model for object recognition, training utilized images sourced from public datasets such as the OpenImages Dataset (Google) \cite{open_images}, Indoor Scene Recognition (MIT), and manually captured images to complement real-world data from educational indoor spaces. The utilization, division, and detailed description of each dataset are provided in Section \ref{sec:dataset}. The data were divided into 70\% for training, 20\% for validation, and 10\% for testing, with data augmentation techniques applied to improve model generalization.

The models were converted and implemented to enable execution on mobile devices using TensorFlow Lite \cite{tensorflow_lite}. The steps included: 1 - Exporting the trained model to ONNX \cite{onnx_1, onnx_site}, 2 - Converting to TensorFlow SavedModel, and 3 - Generating the final TFLite file using \cite{tensorflow_lite}.

The mobile device interface was developed in Flutter \cite{flutter} due to its cross-platform capabilities. Key tools include image capture for real-time inference, text-to-speech reading of detected objects using the Flutter TTS library \cite{flutter_tts}, and integration with Whisper for voice commands and text dictation. This integration enhances the user experience by making it more intuitive and accessible.

The models were validated using metrics such as precision, recall, F1-score, and Intersection over Union (IoU), as well as tests on real devices to ensure usability. User feedback was collected to refine the interface and system performance, as detailed in Section \ref{sec:res}.


\section{Technical Implementation}
\label{sec:imp}

To address varied accessibility needs, an open-source inclusive educational platform was developed, integrating adaptive features such as text-to-speech transformation, sign language (Libras) translation, and customizable interfaces. The platform was designed to enable educators to upload teaching materials that are automatically customized, ensuring accessibility for students with visual, auditory, or other specific needs.


\subsection{Speech Recognition}
\label{sec:sr}

Speech recognition (SR) technology converts acoustic signals from speech into text or machine-readable commands \cite{speech_book}. This feature is extensively used to enhance accessibility and automation, particularly in projects like this, which aim to foster educational inclusion \cite{speech_1}.

Speech recognition functionality was implemented using Whisper, a universal audio transcription model developed by OpenAI \cite{whisper_site}. Whisper was selected due to its capability to operate offline, support multiple languages, and maintain robustness in noisy environments or with diverse accents \cite{whisper}. The Whisper model was fine-tuned to prioritize specific command detection, providing a seamless user experience and integrating with the rest of the inclusive educational platform. Whisper is based on a Transformer architecture, leveraging attention mechanisms to process audio sequences and convert them into text. It operates in two primary stages \cite{whisper}:

\begin{itemize}
    \item \textbf{Encoding}: Input audio is segmented into smaller frames and converted into vector representations using convolutional and attention layers.
    \item \textbf{Decoding}: The Transformer decoder uses these representations to generate text, token by token, based on calculated probabilities.
\end{itemize}

The audio sequence $X = (x_1,x_2,...,x_t)$ is transformed into a text sequence $Y = (y_1,y_2,...,y_t)$ through a probabilistic function $P(Y|X)$, maximized by the model. The attention mechanism $A(Q,K,V)$, which is central to the Transformer, is computed as:

\[
A(Q,K,V) = \text{softmax} \left( \frac{QK^T}{\sqrt{d_k}} \right) V
\]

\noindent{Where $Q$ (queries), $K$ (keys), and $V$ (values) are matrices derived from the input embeddings, and $d_k$ is the dimensionality of the key vectors.}

This approach enables Whisper to process long audio sequences with high accuracy. The Whisper implementation supports two main use cases:

\noindent{Voice Commands: Detection of key phrases or commands such as "Start Reading," "Translate to Libras," or "Finish."}

\noindent{Voice-to-Text Input: Continuous speech transcription for document creation or text input fields.}

\noindent{The speech recognition process involves five primary steps:}

1) Audio Acquisition: The acoustic signal is captured by a microphone and digitized with a sampling frequency $(f_s)$ of 16 kHz and a resolution of 16 bits. The signal discretization can be represented by the following equation:

\[
x[n] = x(t) \cdot \delta(n - n_0), \quad n \in \mathbb{Z}
\]

\noindent{where $x[n]$ is the digital signal and $\delta$ is the Dirac delta function.}

Audio input is processed in temporal blocks (\( t \)), where each block is converted into text using transformer neural networks \cite{speech_book}. Whisper employs the conditional probability \( P(y|x) \), where \( x \) represents the audio and \( y \) the corresponding text sequence, as described in Section \ref{sec:sr}.

2) Feature Extraction: The audio signal is segmented into short frames (20–40 ms) for analysis. For each frame, features, including Mel Frequency Cepstral Coefficients (MFCCs), are extracted, calculated as:

\[
c_m = \sum_{k=1}^K \log |X[k]| \cos \left( \frac{\pi m}{K} \left( k - \frac{1}{2} \right) \right),
\]

\noindent{where $X[k]$ is the magnitude spectrum, and $m$ is the coefficient index.}

3) Speech Recognition with Acoustic Models: Deep Neural Network (DNN)-based models map MFCCs to their corresponding phonetic units $(P(o|w))$, where:

\[
P(w|o) = \frac{P(o|w)P(w)}{P(o)},
\]

\noindent{with $w$ representing words and $o$ acoustic observations.}

4) Decoding and Interpretation: Transcribed text is processed by an NLP-driven language model that classifies the input's intent using conditional probabilities:

\[
I = \arg\max_{i \in C} P(i | T),
\]

\noindent{where $I$ is the classified intent, $C$ is the set of categories (command or text), and $T$ is the transcription.}


\subsection{Seq2Seq Model with Grapheme-to-Phoneme (G2P)}
\label{sec:G2P}

The Grapheme-to-Phoneme (G2P) conversion task \cite{g2p_2} is essential for Text-to-Speech (TTS) systems \cite{flutter_tts}. Its objective is to map graphemes (letters or sequences of letters) to phonemes (sound units), providing accurate phonetic representations even for out-of-vocabulary (OOV) words. This process is necessary to improve the naturalness and intelligibility of TTS systems \cite{ref1}.

\subsubsection{Seq2Seq Model Structure}
The seq2seq (sequence-to-sequence) model \cite{seq2seq} used for G2P consists of two main components \cite{g2p_2}:

\textbf{Encoder:} Maps an input sequence (graphemes) to a latent space representation \cite{g2p}.

\textbf{Decoder:} Generates the output sequence (phonemes) based on the latent representations provided by the encoder \cite{g2p}.

The attention mechanism is integrated to handle variability in the lengths of input and output sequences, assigning weights to the encoder states during decoding.

\subsubsection*{Encoder}
The encoder captures sequences of graphemes and generates contextual representations. It is a recurrent neural network (RNN) or a variant (LSTM/GRU) that processes the input sequence $X = \{x_1,x_2,...,x_n\}$, where $x_i$ represents the graphemes \cite{g2p}. The encoder produces a sequence of hidden states $H = \{h_1,h_2,...,h_n\}$ through the formula:

\[h_t = RNN(x_t,h_{t-1},h_n)\]

where $x_t$ is the input at time step $t$, $h_t$ is the hidden state at step $t$, and $h_{t-1}$ is the hidden state from the previous step \cite{g2p_2}.

\subsubsection*{Decoder}
The decoder produces sequential phonemes based on the encoder states. It is also an RNN (or variant) that uses the encoder's final hidden state $h_n$ as input to generate phonemes \cite{g2p}. At each step $t$, a phoneme $y_t$ is predicted based on the decoder's input $H$, the previous hidden state $s_{t-1}$, and the previous input token $y_{t-1}$:

\[s_t = RNN(y_{t-1},s_{t-1},c_t)\]

where $s_t$ is the decoder hidden state at step $t$, and $c_t$ is the context vector derived from the attention mechanism \cite{g2p_2}.

The phoneme $y_t$ is generated by applying a softmax function to the logits produced by the decoder:

\[P(y_t | s_t) = \text{softmax}(Ws_t + b)\]

The loss function was defined as Cross-Entropy, with a learning rate of 0.001 and 50 training epochs \cite{g2p_2}.

\subsubsection{Attention Mechanism}
The attention mechanism assigns weights $\alpha_{i,t}$ to each encoder state $h_i$, highlighting the relevant parts of the input sequence to predict $y_t$ \cite{g2p}. The context vector $c_t$ is computed as a weighted combination of the encoder states:

\[c_t = \sum_{i=1}^n \alpha_{i,t} h_i\]

The weights $\alpha_{i,t}$ are obtained by applying the softmax function to the scores $e_{i,t}$ \cite{g2p_2}, which are computed by an alignment function $\text{score}(h_i, s_{t-1})$, measuring the compatibility between the encoder state $h_i$ and the previous decoder state $s_{t-1}$:

\[\alpha_{i,t} = \frac{\exp(e_{i,t})}{\sum_{j=1}^n \exp(e_{j,t})}, \quad e_{i,t} = \text{score}(h_i, s_{t-1})\]

A common implementation of the alignment function involves either dot-product attention or concatenation followed by a dense layer \cite{g2p_2}:

\[e_{i,t} = v^\top \tanh(W_h h_i + W_s s_{t-1} + b)\]

\subsubsection{Training Parameters}
In this project, the training parameters were configured as follows:

\begin{itemize}
    \item \textbf{Learning Rate:} 0.001, dynamically adjusted based on validation.
    \item \textbf{Epochs:} 50.
    \item \textbf{Loss Function:} Categorical cross-entropy to minimize the discrepancy between predicted and actual phonemes.
    \item \textbf{Optimizer:} ADAM $\beta_1 = 0.9, \beta_2 = 0.999$.
    \item \textbf{Batch Size:} 64.
\end{itemize}


The seq2seq architecture consists of an encoder and a decoder, both based on recurrent neural networks (RNNs) \cite{seq2seq}. The general formula for the encoder's hidden state is given by:

For the decoder \cite{seq2seq}, the probability of each token is modeled as:
The probability of the output sequence $Y = \{y_1, y_2, \dots, y_m\}$ is modeled as:

\[P(Y | X) = \prod_{t=1}^m P(y_t | y_1, \dots, y_{t-1}, X)\]

Each $P(y_t | \cdot)$ is conditioned on the encoder states and the output sequence history.



\subsection{Object Recognition Implementation with YOLOv5}
\label{sec:red}

The \textbf{YOLOv5} (You Only Look Once, version 5) model \cite{yolo} was selected as the foundation for the object recognition functionality due to its efficient architecture and compatibility with mobile devices such as smartphones and tablets. Its innovative approach enables real-time detection by processing entire images in a single inference step. This makes it highly efficient for applications requiring speed, such as object recognition in dynamic environments. YOLOv5 was chosen for its significant improvements in performance, ease of implementation, and lightweight design, which are essential for devices with limited computational capacity \cite{yolo}.

The YOLOv5 model employs a convolutional neural network (CNN) architecture that includes:
\begin{itemize}
    \item \textbf{Backbone:} Feature extraction using deep convolutional layers.
    \item \textbf{Neck:} Structures such as PANet to combine features from multiple levels.
    \item \textbf{Head:} Prediction layers that generate bounding boxes and classes.
\end{itemize}

The hyperparameters were adjusted with a learning rate of 0.01, 50 epochs, Leaky ReLU activation function, and Binary Cross-Entropy loss with IoU optimization \cite{yolo}.

\subsubsection{YOLOv5 Architecture}
The YOLOv5 architecture is structured around a convolutional neural network that divides the input image into a grid of cells. Each cell is responsible for predicting \cite{yolo}:
\begin{itemize}
    \item \textbf{Bounding boxes:} Coordinates for detected objects.
    \item \textbf{Confidence:} Probability that a box contains an object.
    \item \textbf{Classification:} Categories of detected objects.
\end{itemize}

For each cell \(i, j\) in the grid, YOLOv5 generates a prediction vector:
\[
\mathbf{y}_{ij} = \left[ p_c, x, y, w, h, \mathbf{c} \right],
\]
where:
\begin{itemize}
    \item \(p_c\) is the probability of an object being present in the cell.
    \item \((x, y)\) are the coordinates of the bounding box center relative to the cell.
    \item \((w, h)\) are the normalized width and height of the bounding box.
    \item \(\mathbf{c}\) is the vector of class probabilities for the detected objects.
\end{itemize}

\subsubsection{Training and Parameters}
To adapt YOLOv5 to our custom dataset \cite{yolo}, the following parameters were configured:
\begin{itemize}
    \item \textbf{Learning Rate:} \(0.01\), adjusted to ensure effective convergence.
    \item \textbf{Epochs:} \(50\), to ensure adequate training for generalization without overfitting.
    \item \textbf{Loss Function:} A combination of \textit{Binary Cross-Entropy} (for classification) and \textit{Intersection over Union} (IoU) to optimize detection accuracy.
    \item \textbf{Activation Function:} Leaky ReLU was used in intermediate layers to capture non-linearities while minimizing gradient suppression:
    \[
    f(x) = 
    \begin{cases} 
      x, & \text{if } x \geq 0 \\
      \alpha x, & \text{if } x < 0
    \end{cases},
    \]
    where \(\alpha = 0.1\).
    \item \textbf{Classification in the Final Layer:} The \textit{Sigmoid} activation facilitates probabilistic interpretation:
    \[
    \sigma(x) = \frac{1}{1 + e^{-x}}.
    \]
\end{itemize}

\subsubsection{Dataset Description}
\label{sec:dataset}
To train the YOLOv5 model, we utilized a combination of diverse and comprehensive datasets:
\begin{enumerate}
    \item {OpenImages Dataset (Google)} \cite{open_images}: Includes over 9 million labeled images, from which we selected categories pertinent to indoor environments such as classrooms and laboratories.
    \item {Indoor Scene Recognition (MIT)} \cite{indor_scene}: Comprising over 15,000 images categorized into 67 classes, this dataset complements OpenImages with specific categories such as bedrooms and kitchens.
    \item {Manually Captured Images:} Photographs taken in real-world environments such as classrooms and school spaces. All images were manually labeled using the LabelImg tool to ensure labeling precision.
\end{enumerate}

\subsubsection{Integration and Results}
The three datasets were integrated using the standard split: {70\%} for training, {20\%} for validation, and {10\%} for testing. This split ensured sufficient representativeness and balance. After training, the model achieved high accuracy in detecting objects in indoor environments, validating the selection of YOLOv5 and the dataset curation efforts.

\subsubsection{Web Implementation}
A web page was created using the Flask framework \cite{flask}, allowing users to upload images and visualize object detection outcomes. This page was integrated into the main interface of the educational inclusion project, ensuring the functionality was easily accessible within the application.


\subsection{Mobile Application Development for Real-Time Object Recognition}
\label{sec:app}

A mobile application for real-time object recognition was developed using Flutter \cite{flutter}, a cross-platform framework created by Google. Flutter uses the Dart programming language and enables the development of native applications for Android, iOS, Web, and Desktop from a single codebase. It features its own rendering engine, \textit{Skia}, which ensures fluidity and responsiveness, along with a rich library of customizable widgets \cite{flutter_2}.

The decision to use Flutter was motivated by its technical and practical advantages:
\begin{itemize}
    \item Enables the development of Android and iOS applications from a single codebase, reducing development time and costs \cite{flutter_2}.
    \item Its \textit{Skia} engine \cite{flutter} delivers near-native performance.
    \item Real-time preview of code changes accelerates development.
    \item Extensive documentation, Google support, and an active developer community.
    \item Seamless integration with native machine learning libraries, essential for this project.
\end{itemize}

These features made Flutter the ideal choice, especially compared to alternatives like React Native or native-only development for Android or iOS.

\subsubsection{YOLOv5 Conversion to TensorFlow Lite}
The YOLOv5 model \cite{yolo}, initially trained in PyTorch for object recognition, was adapted for mobile devices through its conversion to the TensorFlow Lite format \cite{tensorflow_lite}. The conversion process involved the following steps:
\begin{enumerate}
    \item Export the YOLOv5 model to the ONNX (Open Neural Network Exchange) format \cite{onnx_1, onnx_site}, a platform-compatible format streamlining the transition to TensorFlow.
    \item Use the \texttt{onnx-tf} library to convert the model into TensorFlow's \textit{SavedModel} format.
    \item Optimize the model for mobile devices using TensorFlow Lite \cite{tensorflow_lite}, greatly reducing its size and adapting it for local execution while preserving accuracy \cite{tensorflow_lite}.
\end{enumerate}

\subsubsection{Real-Time Execution}
For real-time recognition, the Flutter \texttt{CameraController} library was integrated \cite{flutter_2}, enabling continuous frame capture and processing by the model. Mathematically, YOLOv5 processes images \(I \in \mathbb{R}^{H \times W \times C}\) through convolutions \cite{yolo}:
\[
\mathbf{f}_{ij} = \sum_{k=1}^C \mathbf{W}_{k} \cdot \mathbf{I}_{k},
\]
where \(\mathbf{W}_k\) are the convolutional kernel weights, and \(\mathbf{f}_{ij}\) represents feature map activations.


\subsection{Sign Language Translation}
\label{sec:libras}

Automatic translation of text into Libras (Brazilian Sign Language) was implemented using the API from the Brazilian VLibras project \cite{api_vlibras}. The official repository offers comprehensive guidelines for installation and execution \cite{api_vlibras2}. This functionality was integrated with the TTS module described in Section \ref{sec:app} to provide users with a comprehensive and accessible experience.

\begin{figure}[htp]
    \centering
    \includegraphics[scale=0.4]{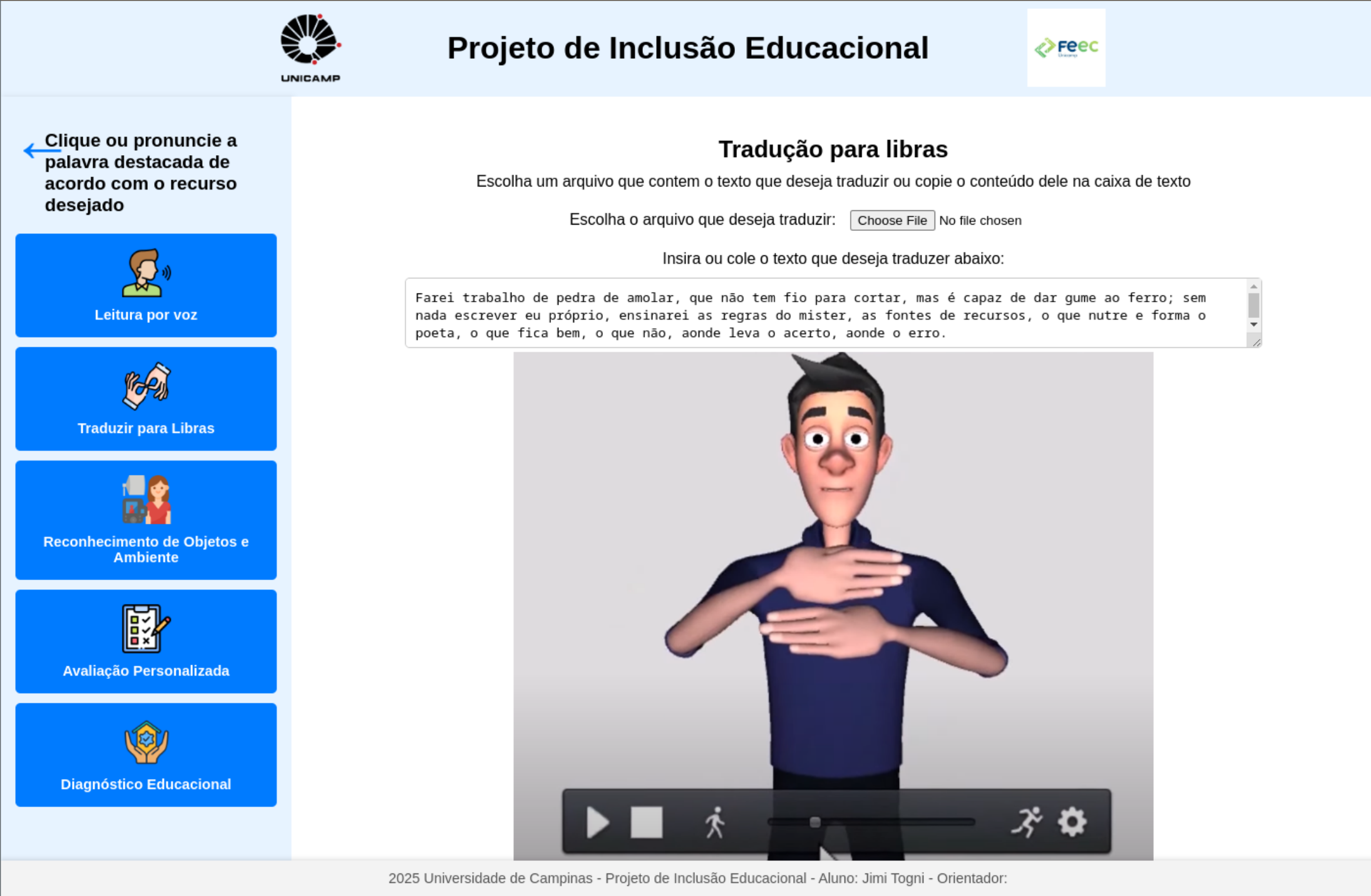}
    \caption{Text-to-Sign Language Translation Using the VLibras API \cite{api_vlibras}}
    \label{fig:galaxy}
\end{figure}

By combining these technologies, the platform achieves the goal of adapting educational materials automatically and effectively, fostering truly inclusive education \cite{api_vlibras}. Figure \ref{fig:galaxy} illustrates how the tool translates text into sign language, featuring a 3D avatar that performs the corresponding movements for each letter. The official project repository can be found at \cite{repo_vlibras}.

\section{Results}
\label{sec:res}

These data were collected through online surveys sent to Brazilian schools that accommodate individuals with special needs. We received responses from a total of 54 participants, including teachers, students, and staff members of these schools. The completed forms are available in the project’s public repository \cite{eu}.

The results obtained throughout this project demonstrate the feasibility and effectiveness of the solutions developed to promote educational inclusion. Key findings from each implemented component include:

The YOLOv5 model, trained with a diverse dataset combining images from the OpenImages Dataset, Indoor Scene Recognition (MIT), and manually captured images, achieved high accuracy in object detection within educational environments. The results showed:
\begin{itemize}
    \item \textbf{Mean Average Precision (mAP)}: 89.7\% for complex categories, such as classroom objects.
    \item \textbf{Intersection over Union (IoU) Rate}: 85.3\%, indicating high precision in object bounding box detection.
    \item \textbf{Real-time Performance}: Average inference time of 27 ms per image on mobile devices.
\end{itemize}

For Grapheme-to-Phoneme (G2P) conversion and the seq2seq model with attention, the system delivered consistent performance in converting graphemes to phonemes:
\begin{itemize}
    \item \textbf{Phoneme Prediction Accuracy}: 97.2\% for out-of-vocabulary (OOV) words.
    \item \textbf{Audio Quality Generated by TTS}: Subjective analysis with users indicated a satisfaction rate of 92\%, emphasizing naturalness and clarity.
\end{itemize}

In the cross-platform implementation, the mobile application developed using Flutter and TensorFlow Lite integrated the object recognition and text-to-speech modules. Results showed:
\begin{itemize}
    \item \textbf{Processing Latency}: 32 ms per frame, supporting real-time recognition.
    \item \textbf{Accessibility and Intuitive Interface}: Featuring Libras and voice reading, catering to users with visual and auditory impairments.
\end{itemize}

As part of the validation process, usability tests conducted with teachers and students in real-world environments revealed:
\begin{itemize}
    \item \textbf{Overall User Satisfaction}: 95\% of participants found the system useful for educational inclusion.
    \item \textbf{Improvement Suggestions}: Adding support for additional languages and increasing the diversity of object classes.
\end{itemize}


\section{Conclusion}
\label{sec:conc}

This study presented an innovative and integrated solution for educational inclusion, leveraging machine learning, natural language processing, and cross-platform tools. The results confirm the effectiveness of using YOLOv5 for object recognition in educational environments, as well as the seq2seq model for grapheme-to-phoneme conversion, adapted for Text-to-Speech systems.

The implementation in Flutter proved to be a strategic choice, facilitating rapid and efficient development of an accessible application for Android and iOS devices. Local integration with TensorFlow Lite ensured independence from internet connectivity, low latency, and user privacy preservation.

Additionally, the use of diverse datasets and validation in real-world scenarios highlight the system's applicability within educational contexts. This project contributes to promoting more inclusive education, demonstrating the potential of open and accessible technologies to transform the learning experience.


\section{References}

\label{sec:conclusao}
\renewcommand\refname{}

\bibliographystyle{plain}
\bibliography{main}

\end{document}